\documentclass[prb,twocolumn,showpacs,superscriptaddress,preprintnumbers,amssymb]{revtex4}
\usepackage{graphicx}
\usepackage{dcolumn}
\usepackage{bm}
\usepackage{wasysym}

\newcommand{\beq}{\begin{equation}}
\newcommand{\eeq}{\end{equation}}
\newcommand{\beqn}{\begin{eqnarray}}
\newcommand{\eeqn}{\end{eqnarray}}

\begin{document}
\title{Gapless Bosonic Excitation without symmetry breaking:
Novel Algebraic Spin liquid with soft Gravitons}
\author{Cenke Xu}
\affiliation{Department of Physics, University of California,
Berkeley, CA 94720}
\date{\today}
\begin{abstract}

A novel quantum ground state of matter is realized in a bosonic
model on three dimensional fcc lattice with emergent low energy
excitations. The novel phase obtained is a stable gapless boson
liquid phase, with algebraic boson density correlations. The
stability of this phase is protected against the instanton effect
and superfluidity by self-duality and large gauge symmetries on
both sides of the duality. The gapless collective excitations of
this phase closely resemble the graviton, although they have a
soft $\omega\sim k^2$ dispersion relation. There are three
branches of gapless excitations in this phase, one of which is
gapless scalar trace mode, the other two have the same
polarization and gauge symmetries as the gravitons. The dynamics
of this novel phase is described by a new set of Maxwell's
equations. The defects carrying gauge charges can drive the system
into the superfluid order when the defects are condensed; also the
topological defects are coupled to the dual gauge field in the
same manner as the charge defects couple to the original gauge
field, after the condensation of the topological defects, the
system is driven into the Mott Insulator phase. In the 2
dimensional case, the gapless soft graviton as well as the
algebraic liquid phase are destroyed by the vertex operators in
the dual theory, and the stripe order is most likely to take place
close to the 2 dimensional quantum critical point at which the
vertex operators are tuned to zero.

\end{abstract}
 \maketitle

\section{Indroduction}

Ground states of quantum many-body systems can behave
qualitatively differently from classical states. Classical ground
states like ferromagnetic order and N\'{e}el order can survive
from quantum fluctuation at zero temperature. The quantum
ferromagnetic state and quantum N\'{e}el ordered state can be
described in a similar way as their classical counterparts.
Searching for nonclassical spin states can be traced back to the
early proposal of the RVB state on frustrated lattices
\cite{anderson1987,zou1987}, which was shown to be of great
importance to the High Tc superconductivity in cuperates.

All the classical orders break certain symmetry, either internal
symmetry or space symmetry. Quantum spin ground states without
classical orders are termed spin liquids. In order to make sure
the quantum ground state has no tendency to order, the low energy
emergent gauge symmetry is usually applied, since as is well
known, the gauge symmetry cannot be broken without condensation of
matter fields \cite{fradkin1978}. After two decades of work,
several types of nonclassical ground states have been identified.
For instance, the spin liquid with $Z_2$ gauge symmetry has been
realized in either quantum dimer model on triangular lattice
\cite{sondhi2000}, or spin-1/2 model on Kagom\'{e} lattice
\cite{balents2002}. The existence of the $Z_2$ spin liquids in 2+1
dimensional space is based on the fact that the quantum $Z_2$
gauge theory has a deconfined phase, which manifests itself as a
disordered spin state with fractionalized spin excitations. In
these models, the ground states have no classical order, i.e.
there is no symmetry breaking. The ground state degeneracy depends
on the topology of the space manifold, which is termed topological
order. The excitations contain the deconfined gapped $Z_2$ gauge
charges and the $Z_2$ vortices (which are usually called visons).

Another type of spin liquids contain gapless collective
excitations (which are different from the gapless Magnon
excitations in spin ordered state), and the spin-spin correlation
functions fall off algebraically, i.e. the state is in a stable
critical phase. These algebraic spin liquids usually involve
$U(1)$ gauge field, or $U(1)$ gauge field interacting with gapless
matter fields. The stability of algebraic spin liquids is very
tricky, because algebraic spin liquids are critical, presumably
there are supposed to be relevant perturbations which can drive
the system into an ordered phase. Actually in most cases the
algebraic liquid phases are fine-tuned, for instance, the
Rokhsar-Kivelson (RK) point of quantum dimer model on square
lattice \cite{rokhsar1988,ashvin2004a}. Due to the presence and
proliferation of monopoles in 2+1 dimensional systems
\cite{polyakov1977,polyakov1987}, the gapless photon excitations
are generally gapped out, and the matter fields are confined. In
the 3+1 dimensional space, the compact gauge theory has a
deconfined photon phase. Based on this result, several photon
liquid phase has been proposed
\cite{sondhi2003,wen2003,hermele2004}. In the 2+1 dimensional
systems, the monopoles are almost always proliferating. The
algebraic liquid phase is very hard to survive. The monopoles are
only irrelevant at certain critical point, for instance the RK
point of the quantum dimer model \cite{rokhsar1988}, as well as
the transition between valence bond solid phase and N\'{e}el state
for spin-1/2 antiferromagnetic system
\cite{senthil2004,ashvin2004}. Recently it has been argued that at
the large $N$ level ($N$ is the number of flavors of matter
fields), there is a stable algebraic liquid phase in 2+1
dimensional space \cite{Hermele2004a}.

Gapless bosonic excitation is one property which characterizes
criticality. Since here the critical phase is stable, it implies
that even without any continuous symmetry breaking there is a
gapless bosonic excitation invulnerable to perturbations. This is
actually a quite amazing property. As is well-known to all, the
Goldstone theorem is one way to protect the gaplessness of bosonic
systems \cite{goldstone}. By breaking continuous global symmetry,
the coset of the unbroken symmetry subgroup corresponds to the
gapless modes, which are called Goldstone modes. Almost all the
stable gapless bosonic excitations are related to certain
continuous symmetry breaking, for instance, the phonon in solids
is related to the spatial translational symmetry breaking, the
magnon is related to the $SU(2)$ spin symmetry breaking. However,
the gapless critical excitations of algebraic spin liquid phase do
not rely on any symmetry breaking. Therefore the algebraic spin
liquid phase kindled the questions like "original lights". It has
been proposed that the photon, which is one of the most
fundamental particles in the universe, might not be so
fundamental, it could be collective excitations of lattice spin
models \cite{wen2003}.

All the stable algebraic liquid phases which have been found so
far involve $U(1)$ gauge field theory. This $U(1)$ gauge symmetry
does not exist in the high energy (microscopic) model, instead it
emerges at low energy Hamiltonian, due to the effective constraint
$\vec{\nabla} \cdot \vec{E} = \rho$ imposed by the spin
interaction. $\rho$ is the background static charge distribution.
Although the microscopic models which have been proposed so far
\cite{hermele2004,wen2003,sondhi2003} are different on the lattice
scale, the 3+1 dimensional photon liquid phases are all the same
at long scale. In one previous piece of work \cite{xu2006} and the
current work, a new type of algebraic spin (boson) liquid phase
has been realized. Based on the standard spin-boson mapping ($S^z
= n - \bar{n}$, and $S^\dagger \sim \exp(i\theta)$), the model is
presented in the bosonic version. The algebraic boson liquid state
studied in this paper is a new state of matter, it broadens the
family of algebraic liquid phases. The new algebraic liquid phase
does not involve $U(1)$ gauge theory. Instead, the gauge symmetry
of this model is identical to the gauge symmetry of linearized
Einstein gravity. There are three branches of gapless collective
excitations in this new bosonic algebraic liquid phase. One of the
collective excitations is a scalar mode, the other two gapless
collective excitations have the same gauge symmetry and
polarizations as the gravitons. However, the graviton excitations
in our theory have a softened quadratic dispersion, $\omega \sim
k^2$. The spin-spin (boson density) correlation functions fall off
algebraically, with an exponent bigger than that of the 3+1
dimensional photon liquid. Like gravitational theory, the basic
variables in this theory are symmetric rank-2 tensor, which in the
gravity language are the linearized metric tensor. The dynamics of
the graviton phase is described by a new set of Maxwell's
equations, with the vectors $\vec{E}$ and $\vec{B}$ replaced by
rank two symmetric tensors. The charges and topological defects
enter the Maxwell's equations in a special form, and the
condensations of charges and topological defects will drive the
liquid phase into the superfluid phase and the Mott insulator
phase respectively.

This paper is organized as follows. In the second section, a brief
review of the 3+1 dimensional photon spin liquid phase is
presented. The discussion of the graviton spin liquid will follow
the same logic as the photon spin liquid phase. In the third
section, the 2+1 dimensional version of the graviton model is
discussed. In the 2+1 dimensional case, the graviton theory is
dual to a scalar boson model with quadratic dispersion. However,
the vertex operators will generally gap out the graviton
excitations. In the forth section, the 3+1 dimensional graviton
model is described, and it is shown that the graviton phase
(Gaussian phase) is self-dual, and hence stable. In section V,
more properties of this model are derived. The boson density
correlation function is calculated, and a new set of Maxwell's
equations are supposed to describe this Gaussian phase. In section
VI, we discuss the possible experimental realization of this
model, as well as the future work. If the model is written in
terms of fermionic operators instead of bosonis operators, novel
non-fermi liquid behavior is expected. It is also possible to
develop this theory to be a new candidate of quantum gravity
theory.

\section{review of 3+1 dimensional photon liquid phase}

\begin{figure}
\includegraphics[width=2.7in]{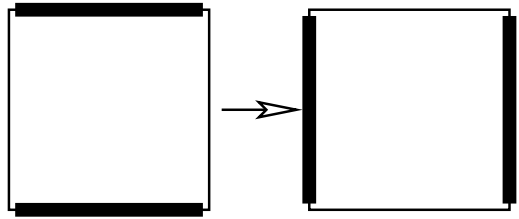}
\caption{The flipping term of the dimer model.} \label{fdimer}
\end{figure}

Let us first briefly review some basic properties of the 3+1
dimensional photon liquid phase as a warmup. Although many
different models have been proposed in the last few years
\cite{sondhi2003,hermele2004,wen2003}, the phases obtained have
very similar properties at long scale. Therefore, let us take the
3 dimensional Quantum Dimer Model on cubic lattice as an example.
On the cubic lattice, every site is shared by six links, and only
one of those links is occupied, by exactly one dimer. On every
square face, if two parallel links are occupied by dimers, they
can be resonated to dimers perpendicular to the original ones
(Fig. \ref{fdimer}). Besides this resonating term, another
diagonal weight term for each flippable plaquette is also included
in the Hamiltonian \beqn H = \sum -t(| \parallel \rangle\langle =
| + h.c.) + V(| \| \rangle\langle \| | + | = \rangle\langle =
|)\label{dimer}.\eeqn

\begin{figure}
\includegraphics[width=2.2in]{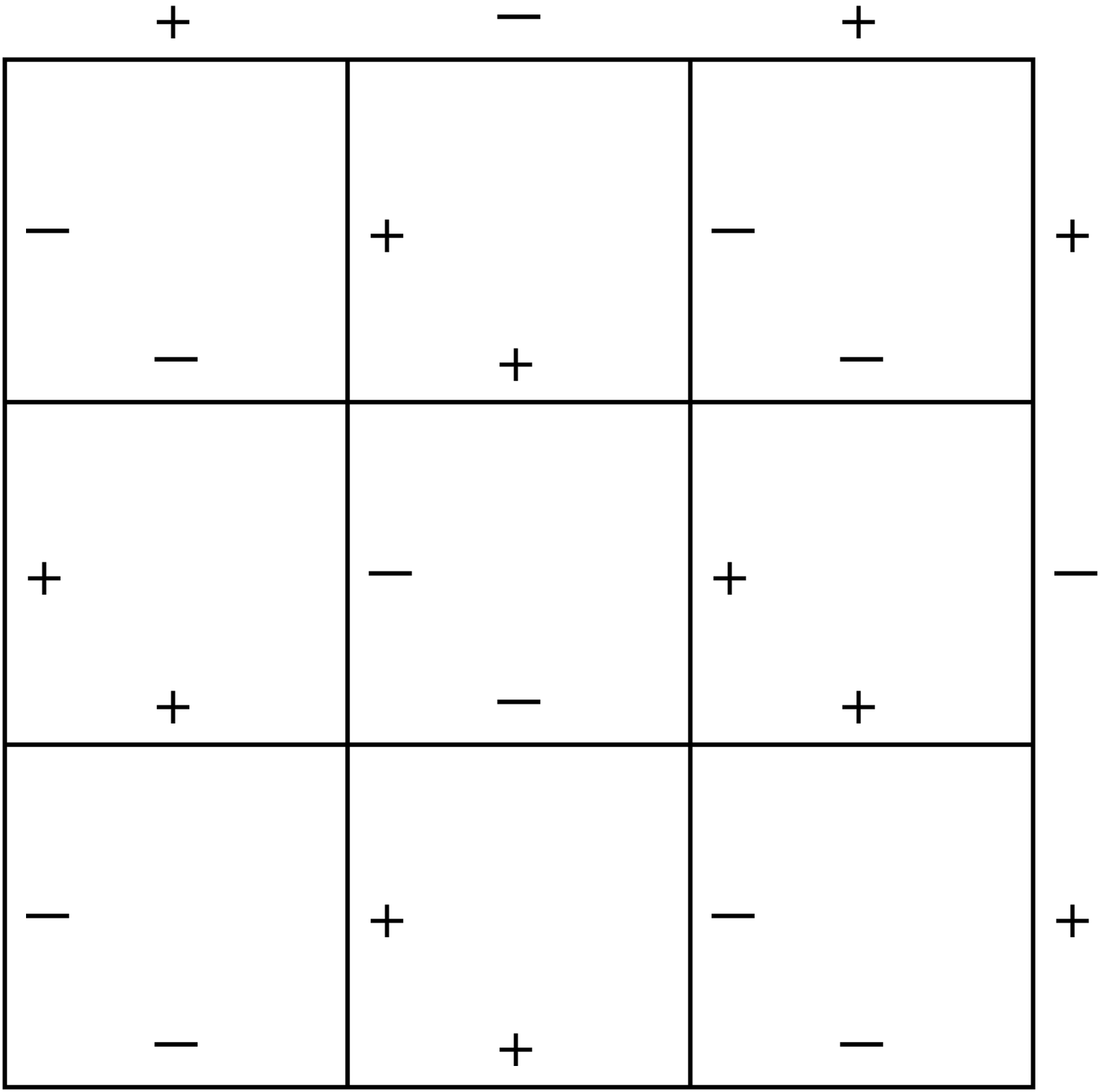}
\caption{The sign convention of mapping from rotor number onto the
vector electric field $\vec{E}$ in XY plane.} \label{signdimer}
\end{figure}

The dimer model can be mapped onto a rotor model. A rotor number
can be defined on each link to describe the presence or absence of
dimers: $n = 1$ when the link is occupied by a dimer, and $n = 0$
when the link is empty. The Hilbert space of this quantum system
is a constrained one, with the constraint $\sum_{i = 1}^6 n_i = 1$
around each site. Let us define quantity $E_{i,\hat{a}}$ as
$E_{i,\hat{a}} = (-1)^{i_x+i_y}n_{i,\hat{a}}$ (The sign
distribution is shown in Fig. \ref{signdimer}). Now, the
constraint of this system can be rewritten as the Gauss's law for
electric fields, $\partial_i E_i = \pm 1$. Notice that because the
quantity $E_{i,\hat{a}}$ is defined on links, a natural vector
notation can be applied: $\vec{E}_i = (E_{i,\hat{x}},
E_{i,\hat{y}},E_{i,\hat{z}})$. Here the derivatives are all
defined on the lattice $\partial_iE_j(x) = E_j(x + \hat{i})-
E_j(x) $.

The background charge distribution $\pm 1$ on the cubic lattice
plays a very important role in the solid phase, i.e. the confined
phase. The form of the crystalline phase can be determined from
the Berry phase induced by the background charge distribution
\cite{ashvin2004,hermele2004}. However, as we are focusing on the
algebraic liquid phase, the background charge is not very
important. Let us instead impose the constraint $\partial_i E_i =
0$. Because of this local constraint, the low energy physics will
be invariant under the gauge transformation $\vec{A} \rightarrow
\vec{A} + \vec{\nabla}f$ ($\vec{A}$ is the conjugate variable of
$\vec{E}$, which is related to the phase angle $\theta$ associated
with each rotor number $n$ through $\vec{A}_i(x,\hat{i}) =
(-1)^{i_x + i_y}\theta_{x,\hat{i}}$), which is exactly the gauge
symmetry of the $U(1)$ gauge theory. Now the effective Hamiltonian
of this theory should be invariant under the gauge transformation.
The Hamiltonian (\ref{dimer}) can be effectively written as \beqn
H = \sum - \tilde{t}\cos(\vec{\nabla}\times\vec{A}) +
1/(2\kappa)\vec{E}^2 \label{photonhamil}\eeqn This Hamiltonian is
the 3 dimensional compact QED, which should have a deconfined
photon phase \cite{polyakov1977,polyakov1987}. In this phase, the
cosine functions in (\ref{photonhamil}) can be expanded, and the
system can be described in the following Gaussian fixed point
Lagrangian \beqn L = (\partial_\tau \vec{A})^2 - c^2(\vec{\nabla}
\times \vec{A})^2. \label{photonaction} \eeqn The constraint
$\vec{\nabla}\cdot \vec{E} = 0$ introduces a Lagrange multiplier
$A_0$ to the Lagrangian $A_0(\nabla_i\cdot E_i) $. After
integrating out $\vec{E}$, the gauge invariance can be enlarged to
3+1 dimensional space time, $A_\mu \rightarrow A_\mu +
\partial_\mu f$, with $\mu = 0, 1, 2, 3$. The 3+1 dimensional
Lagrangian reads \beqn L \sim - F_{\mu\nu}F^{\mu\nu}. \eeqn

The effective Lagrangian for the photon phase has been derived
above, and the reason for the existence of the photon phase is
actually the remarkable self-duality of this photon phase and the
gauge symmetry, as discussed below.

First, because the constraint $\partial_i E_i = 0$ is strictly
imposed on this system, the matter field, i.e. the defect which
violates this constraint, is absent. As is well known, the $U(1)$
gauge symmetry can be spontaneously broken due to the condensation
of the matter fields. However, without matter field, the local
gauge symmetry cannot be broken spontaneously \cite{fradkin1978}.
The superfluid phase of the original system implies $\langle
\theta \rangle \neq 0$, written in the low energy variables, it
reads $\langle \vec{A}\rangle \neq 0$. The nonzero expectation of
vector potential $\vec{A}$ breaks the local gauge symmetry.
Therefore the dimer superfluid order is ruled out.

The other possible instability of the photon phase is towards the
gapped solid phase, which can be analyzed in the dual theory. We
can introduce the dual vector $\vec{h}$ and dual momentum vector
$\vec{\pi}$ ($\vec{h}$ and $\vec{\pi}$ are both defined on the
faces of the cubic lattice) as \beqn \vec{E} = \vec{\nabla}\times
\vec{h}, \vec{\nabla}\times \vec{A} = \vec{\pi}.
\label{dualh}\eeqn. One can check the commutation relation and see
that $\vec{h}$ and $\vec{\pi}$ are a pair of conjugate variables.
The Gauss's law constraint on this system is automatically solved
by the vector $\vec{h}$. Now, the field theory for the photon
phase is self-dual: \beqn L = (\partial_\tau \vec{A})^2 -
c^2(\vec{\nabla}\times \vec{A})^2, \vec{\nabla}\cdot\vec{E} =
0,\cr\cr L = (\partial_\tau \vec{h})^2 - c^2(\vec{\nabla}\times
\vec{h})^2, \vec{\nabla}\cdot\vec{\pi} = 0. \label{photonself}
\eeqn

The violation of the dimer constraint (for instance the hole of
the doped dimer model) can be viewed as the gauge charges. The
defects couple to the gauge field vector in a gauge invariant
manner \beqn L_e = -t\cos(\vec{\nabla} \theta^{(e)} - \vec{A}) +
\cdots \label{matter}.\eeqn $\theta^{(e)}$ is the phase angle of
the defect creation operator, which plays the role of the electric
charge in the QED language. There are usually two flavors of
matter fields, since besides the $U(1)$ gauge symmetry, there is
an extra global $U(1)$ symmetry, which corresponds to the global
conservation of total holon number. When the defects condense, the
gauge bosons are gapped out by Higgs mechanism, the system enters
the superfluid order, and the global $U(1)$ symmetry is
spontaneously broken and becomes the gapless Goldstone phason mode
in the superfluid phase \cite{balents2005a,balents2005b}.

In principle, a vertex operator $\cos(2\pi N\vec{h})$ is supposed
to exist in the dual Hamiltonian, due to the fact that $\vec{E}$
only takes on integer values. $N$ is an integer depending on the
Berry phase of the vertex operator. When this vertex operator is
relevant, it will gap out the photon excitation and drive the
system into a crystalline phase, according to the Berry phase.
However, the vertex operator is irrelevant in the photon phase.
Notice that, because the theory is self-dual, the dual theory has
the same gauge invariance $\vec{h} \rightarrow \vec{h} +
\vec{\nabla} f$ as the original theory. Also, vector $\vec{\pi}$
is subject to the same constraint as $\vec{E}$, $\partial_i\pi_i =
0$ (\ref{photonself}). However, the vertex operator $\cos(2\pi
N\vec{h})$ breaks the gauge symmetry and thus the correlation
function between two vertex operators is zero at the Gaussian
fixed point, i.e. the vertex operator is irrelevant in this
Gaussian theory.

Because the theory is self-dual at the Gaussian phase, the
magnetic monopoles (the dual charges) should couple to the dual
vector potential $\vec{h}$ in the same manner as the coupling
between electric charge and original vector potential $\vec{A}$
(\ref{matter}) \beqn L_m = -t^\prime\cos(2\pi
N\vec{\nabla}\theta^{(m)} - 2\pi N\vec{h}). \eeqn $N$ is again
introduced by the Berry phase, corresponding to the multi-monopole
event. Unless the dual charges (the monopole) condense and break
the dual $U(1)$ gauge symmetry, the photon phase is always stable.
The gaplessness of this phase is protected by both the
self-duality and the gauge symmetry.

The photon phase is an algebraic liquid phase, the dimer
density-density correlation function falls off algebraically. When
$t = V$ in equation (\ref{dimer}), the lattice model can be solved
exactly, and the equal-time density correlation falls off as
\cite{sondhi2003} \beqn\langle n(0)n(r) \rangle \sim 1/r^3.\eeqn

\section{graviton model in 2 dimensional space}

\begin{figure}
\includegraphics[width=3.0in]{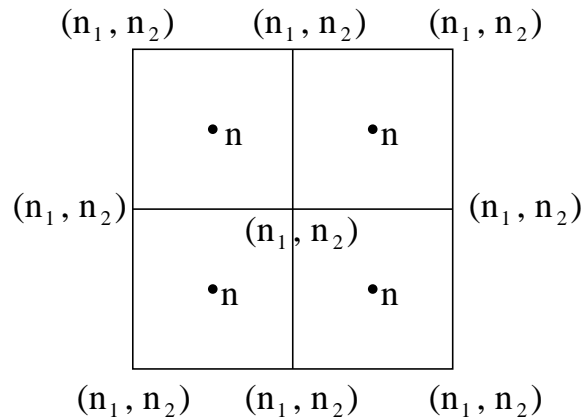}
\caption{The structure of the 2d lattice. on each site there are
two orbital levels, the occupation number is $(n_1, n_2)$. On each
face center there is one orbital level, with occupation number
$n$.} \label{2dlattice}
\end{figure}

Let us begin with the 2 dimensional example. This 2 dimensional
model is built on a square lattice with quantities defined on both
sites and centers of plaquettes. Let us assume there is one
orbital level on the center of each plaquette, and two orbital
levels on each site (see Fig.\ref{2dlattice}). The Hamiltonian of
the system contains 3 terms, $H = H_0 + H_1 + H_2$. $H_1$ is
merely the nearest neighbor hopping term between the sites and the
centers of plaquettes \beqn H_1 = \sum_{<i,\bar{j}>}\sum_{a = 1}^2
-t(b^\dagger_{a,i}b_{\bar{j}} + h.c.).\label{2dh1}\eeqn Here $i$
denotes the site of the lattice, and $\bar{j}$ denotes the center
of plaquette, the summation is over all the hoppings between each
site and its 4 nearest neighbor plaquettes. $H_2$ is an on-site
interaction $H_2 = U(n - \bar{n})^2$, which fixes the average
filling per orbital state. $\bar{n}$ is the average particle
number per site, for simplicity it is taken to be $1$.

The most important term in this Hamiltonian is $H_0$. It is a two
body interaction between particle numbers with a special form.
Each link of the square lattice is shared by two plaquettes and
two sites. We denote the link between sites $i$ and $i+\hat{x}$ as
$(i,\hat{x})$, and denote the two plaquettes connecting to the
link as $\bar{i} = i + 1/2 \hat{x} + 1/2 \hat{y}$ and $\bar{i} -
\hat{y} = i- 1/2 \hat{y} + 1/2 \hat{x}$. The term in $H_0$ which
involves this link reads \beqn V( n_{i + 1/2 \hat{x} + 1/2
\hat{y}} + n_{i- 1/2 \hat{y} + 1/2 \hat{x}} + 2n_{1,i} +
2n_{1,i+\hat{x}} - 6\bar{n})^2, \cr \label{consx}\eeqn and the
interaction term in $H_0$ involving the link $(i,\hat{y})$ is
\beqn V( n_{i + 1/2\hat{x} + \hat{y}} + n_{i- 1/2 \hat{x} + 1/2
\hat{y}} + 2n_{2,i} + 2n_{2,i+\hat{y}} - 6\bar{n})^2. \cr
\label{consy}\eeqn Notice that, for links in $\hat{x}$ direction
only $n_{1}$ is in this interaction term, and for links in
$\hat{y}$ direction only $n_2$ is involved. If $V$ is much bigger
than $t$, the summation in the brackets in both (\ref{consx}) and
(\ref{consy}) should be zero, this becomes a constraint on the
Hilbert space if $V$ is large.

\begin{figure}
\includegraphics[width=2.2in]{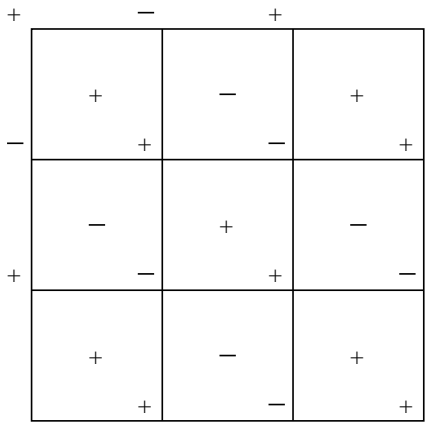}
\caption{The sign convention in the definition of symmetric tensor
$E_{ij}$. After introducing the signs on the lattice, the
constraint imposed by $H_0$ can be written compactly as
(\ref{cons}).} \label{signgrav}
\end{figure}

Because the square lattice is a bipartite lattice, we can stagger
the sign for each sublattice. Let us define new variables
$E_{xx}(i) = \eta_{i} (n_{1,i} - 1)$, $E_{yy}(i) = \eta_i
(n_{2,i}-1)$; and $E_{xy}(\bar{i}) = \eta_{\bar{i}}
(n_{\bar{i}}-1)$. $\eta$s are signs defined on sites and centers
of plaquettes, $\eta_i (\eta_{\bar{i}}) = \pm 1$, with $+$ for
sublattice $A$ ($\bar{A}$) and $-$ for sublattice $B$ ($\bar{B}$).
The constraint approximately imposed by $H_0$ can now be rewritten
as \beqn 2\partial_xE_{xx} +
\partial_yE_{xy} = 0,\cr\cr \partial_{x}E_{xy} + 2\partial_yE_{yy}
= 0.\label{cons}\eeqn Again all the derivatives are defined on
lattice. The convention of the staggered signs is depicted in Fig.
\ref{signgrav}.

The nearest neighbor hopping $t$ term will generate certain "ring
exchange" term  by perturbation theory, which is allowed by the
constraint (\ref{consx}) and (\ref{consy}) at low energy. However,
without doing the perturbation literally, one can guess the form
of the ring exchange term from the form of the constraint
(\ref{cons}). Just like the constraint $\partial_iE_i = 0$
generates gauge transformation $A_i \rightarrow A_i +
\partial_i\varphi$, the current constraint on $E_{ij}$
(\ref{cons}) will generate gauge transformation for its conjugate
variable $A_{ij}$ \beq A_{ij} \rightarrow A_{ij} + \partial_if_j +
\partial_jf_i, \label{gauge}\eeq and the low energy ring exchange term generated
from perturbation theory should be invariant under this gauge
transformation. $A_{ij}$ is related to the phase angles of the
original boson creation and annihilation operators by introducing
staggered sign distribution $\eta_i$: $A_{xx}(i) =
\eta_i\theta_{1,i}$, $A_{yy}(i) = \eta_i\theta_{2,i}$ and
$A_{xy}(\bar{i}) = \eta_{\bar{i}}\theta_{\bar{i}}$.

One may have already noticed that, the gauge transformation
(\ref{gauge}) is exactly the gauge transformation for the graviton
if we view rank two tensor $A_{ij}$ as the linearized metric
tensor on 2 dimensional space. Then the only gauge invariant ring
exchange is the curvature tensor, which is
$R_{\alpha\mu\beta\nu}$. Written in terms of linearized metric
tensor, the curvature tensor is \cite{wheeler1973} \beqn
R_{\alpha\mu\beta\nu} = 1/2(A_{\alpha\nu,\mu\beta} +
A_{\mu\beta,\alpha\nu} - A_{\mu\nu,\alpha\beta} -
A_{\alpha\beta,\mu\nu}).\cr\eeqn Here the notation in general
relativity has been used, $A_{\alpha\nu,\mu\beta} =
\partial_{\mu}\partial_{\beta} A_{\alpha\nu}$.

\begin{figure}
\includegraphics[width=3.2in]{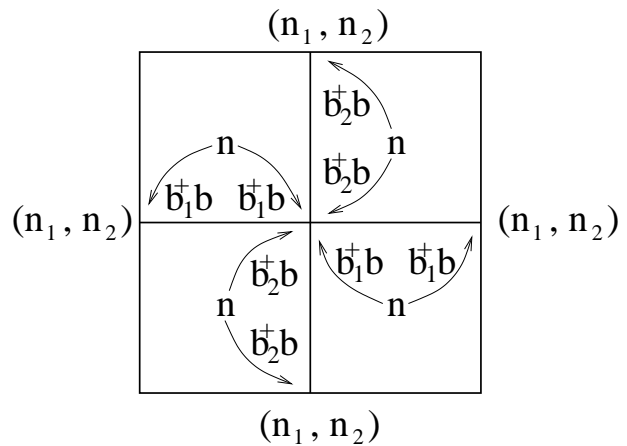}
\caption{The ring exchange of the 2 dimensional lattice.
Elementary hoppings are between site and nearest neighbor face
center. Gauge invariant ring exchange term can be represented as
$\cos(R_{xyxy})$, it involves eight single hoppings of bosons.}
\label{2dring}
\end{figure}

In 2 dimensional space, although the curvature tensor has four
indices, there is only one nonzero independent component, which is
\beqn R_{xyxy} = \frac{1}{2}(A_{xx,yy} + A_{yy,xx} - 2A_{xy,xy}).
\eeqn At eighth order perturbation of $t$, a ring exchange term
$\cos(R_{xyxy})$ is generated. Now the low energy effective
Hamiltonian reads \beqn H_{eff} = -\tilde{t}\cos(R_{xyxy}) +
\frac{1}{2\kappa}(E^2_{xx} + E^2_{yy} + a E^2_{xy})
\label{2dhamil}\eeqn The cosine term in this Hamiltonian is ring
exchange, $\tilde{t} \sim t^8/V^7$. If the original boson language
is taken, one of the ring exchanges is shown in Fig.\ref{2dring},
one can clearly see that the ring exchange term involves 8
independent nearest neighbor hoppings, this term only takes place
at the eighth order perturbation of $t$.

The first conclusion drawn from the gauge symmetry (\ref{gauge})
is that, this system cannot be in the superfluid phase. Because
nonzero expectation value of the boson creation operator implies
nonzero expectation of $A_{ij}$. Any linear combination of
$A_{ij}$ will break the local gauge symmetry, which is not allowed
without matter field \cite{fradkin1978}. Another way to view this
point is through the Hamiltonian. The hopping term $ H_1$ in
(\ref{2dh1}) has global $U(1)$ symmetry, which corresponds to the
global conservation of total boson number. One might wonder
whether this global $U(1)$ symmetry can be spontaneously broken,
i.e. the system becomes superfluid. However, $H_0$ opens a very
big energy gap to excitations that do not conserve boson number.
The big charge gap rules out the possibility of superfluidity.
Because of the graviton gauge symmetry, the polarization of the
collective excitation should automatically be the same as the
gravitons. However, one crucial difference from the photon liquid
phase is that, here the curvature tensor is the second order
derivative of $A_{ij}$, if there is a phase in which we can expand
the cosine functions in (\ref{2dhamil}), i.e. there is a Gaussian
phase or Gaussian fixed point, the dispersion relation of the
gapless collective modes should be $\omega \sim k^2$.
Unfortunately, just like the monopole proliferation in 2+1
dimensional QED, the topological defects in the current case also
generally proliferate, and will gap out the graviton excitations.

The effect of the topological defects can be described in the dual
formalism. Dual variable $\pi$ and $h$ can be defined as follows
\beqn E_{xx} = \partial^2_{y}h, E_{yy} =
\partial^2_{x}h, \cr\cr E_{xy} = -2\partial_{x}\partial_{y}h,
2R_{xyxy} = \pi. \label{definedual}\eeqn $h$ and $\pi$ are
quantities defined on lattice $i$. One can check the commutator
and see $h$ and $\pi$ are a pair of conjugate variables,i.e.
$[h_i,\pi_j] = i\delta_{ij}$. Because the definition of the dual
variable $h$ only involves the second derivatives, the dual
Lagrangian should be invariant under following transformation
\beqn h \rightarrow h + A x + B y + C \label{dualsymm}\eeqn $A$,
$B$, $C$ are arbitrary constant integers.

\begin{figure}
\includegraphics[width=2.0in]{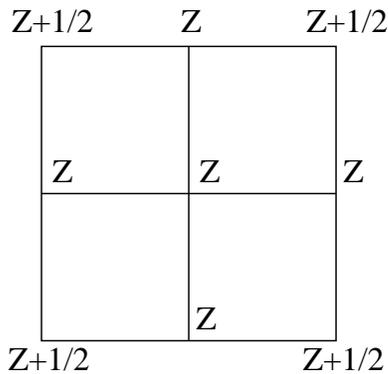}
\caption{The distribution of the dual variable $h$. $h$ cannot all
be integers over the whole 2 dimensional plane, instead, one
quarter of $h$ have to be half-integers.} \label{2dh}
\end{figure}

The dual Lagrangian now reads \beqn L_{dual} = (\partial_\tau h)^2
- \rho_4 h(\partial^4_x + \partial^4_y + 4a
\partial^2_x\partial^2_y)h + \cdots \cr \label{dualac}\eeqn The
ellipses include vertex operators in this dual theory. Notice
that, although $E_{ij}$ are all integers, the representation of
$E_{xy}$ in (\ref{definedual}) contains factor 2 on the definition
of the dual variables, thus $h$ on some lattice sites have to be
half integers, and the Berry phase will cause oscillation of the
signs of the vertex operators on the lattice space. The
distribution of $h$ is shown in Fig. \ref{2dh}. The leading
unoscillating vertex operators are \beqn L_{vertex} = - \alpha
\cos(4\pi h)- \gamma (\cos(4\pi\partial_xh) + \cos(4\pi\partial_y
h)).\cr \label{2dvertex}\eeqn

This Lagrangian (\ref{dualac}) looks exactly like the Lagrangian
at the Rokhsar-Kivelson point for 2 dimensional dimer model on
square lattice \cite{henley1997, fradkin2004} except for the
vertex operators. The leading kinetic term of this Lagrangian is
proportional to $k^4$. The $k^2$ term is ruled out by the symmetry
(\ref{dualsymm}). Since $k^2$ term does not exist in the dual
Lagrangian (\ref{dualac}), as long as $\gamma$ is tuned to zero
and $\rho_4$ is smaller than a critical value, the vertex operator
is irrelevant, and the system is a liquid phase without any order
\cite{fradkin2004}. However, the vertex operator proportional to
$\gamma$ is relevant generically (and hence $\alpha$ term is
dangerously irrelevant). The relevant vertex operator will drive
the system into crystalline phase. The crystalline pattern close
to the critical point with $\gamma = 0$ can be predicted from the
field theory. The representations of density operators in terms of
the low energy field variable are \beqn n_1,n_2 -1 \sim
(-1)^x\sin(2\pi\partial_xh) + (-1)^y\sin(2\pi\partial_yh),\cr\cr n
- 1\sim - (-1)^x\sin(2\pi\partial_xh) -
(-1)^y\sin(2\pi\partial_yh),\cr\cr \cos(R_{xyxy}) \sim
(-1)^y\cos(2\pi\partial_xh) + (-1)^x\cos(2\pi\partial_yh).\cr\eeqn
When the vertex operators in (\ref{2dvertex}) are relevant, the
order parameters above will take nonzero expectation values. Thus
the $\gamma$ operator tends to drive the system into the $(\pi,0)
+ (0,\pi)$ state for either particle density or plaquette density,
depending on the sign of $\gamma$. Each case has 4 degenerate
ground states.

\section{graviton model in 3 dimensional space}

\begin{figure}
\includegraphics[width=2.8in]{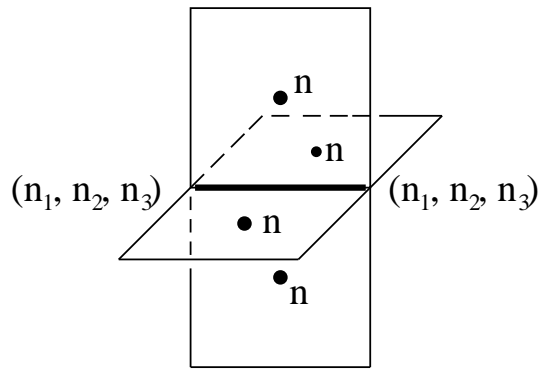}
\caption{The distribution of boson numbers on fcc lattice. The
link (in bold font) is shared by four plaquettes and two sites.
Every site is occupied by three orbital levels, and on every
plaquette there is one orbital level.} \label{3dlattice}
\end{figure}

The model in 3 dimensional space is defined on the fcc lattice,
since physical quantities are defined on both sites and the
centers of each plaquette. Let us assume there are 3 orbital
levels on each site, and 1 orbital level on each face center. The
particle number on the face center is denoted as $n$, and the
particle numbers on sites are denoted as $(n_1,n_2,n_3)$
(Fig.\ref{3dlattice}).

The Hamiltonian for this system still contains three parts, $H =
H_0 + H_1 + H_2$. $H_1$ is the nearest neighbor hopping between
sites and their nearest face centers, and also between adjacent
face centers (notice that the adjacent face centers have the same
distance as the site and its nearest face center). \beqn H_1 =
\sum_{<i,\bar{j}>}\sum_{a = 1}^3-tb^\dagger_{a,i}b_{\bar{j}} -
\sum_{<\bar{i},\bar{j}>}t b^\dagger_{\bar{i}}b_{\bar{j}}+h.c.
\label{3dh1}\eeqn, $H_2$ is the on site interaction $H_2 = U(n -
1)^2$, which fixes the average filling of the fcc lattice. $H_0$
is the interaction term involving links in all 3 directions. For
example, for the link in $(i,\hat{x})$, the interaction term reads
\beqn H_0 = V(n_{i + 1/2\hat{x} + 1/2\hat{z}} + n_{i + 1/2\hat{x}
- 1/2\hat{z}} + n_{i + 1/2\hat{x} + 1/2\hat{y}} \cr\cr + n_{i +
1/2\hat{x} - 1/2\hat{y}} + 2n_{1,i} + 2n_{1,i+\hat{x}} - 8)^2.
\cr\label{h0}\eeqn. The links in $\hat{y}$ and $\hat{z}$
directions are treated similarly. If the bracket in equation
(\ref{h0}) is expanded, it becomes the usual two body repulsion
term.

When $H_0$ becomes the dominant term in the Hamiltonian, it
effectively imposes a constraint on the system. Again the best way
to view this constraint is by introducing a staggered sign and
defining new variables, similar to the electric field in the dimer
model discussed before. Let us define a rank-2 tensor $E_{ij}$.
The off-diagonal terms are defined on face centers as $E_{ij} =
\eta_r (n - 1)$. $n$ is located at one of the $\hat{i}\hat{j}$
face centers; the diagonal term is defined on sites as $E_{ii} =
\eta_r (n_{i} - 1)$ with $i = 1,2,3$. $\eta_r = \pm 1$ and the
distribution of sign $\eta_r$ is shown in Fig. \ref{signgrav}

After introducing the sign $\eta$, the constraint effectively
imposed by (\ref{h0}) can be compactly written as \beqn
2\partial_{x}E_{xx} + \partial_y E_{xy} +
\partial_{z}E_{xz} = 0, \cr \partial_{x}E_{xy} + 2\partial_y E_{yy} +
\partial_{z}E_{yz} = 0, \cr \partial_{x}E_{xz} + \partial_y E_{yz} +
2\partial_{z}E_{zz} = 0.\label{cons3}\eeqn A violation of this
constraint can be interpreted as a charged defect excitation and
can drive the system into an ordered boson superfluid state after
condensation. We will discuss the transition in section V. In the
current section we only focus on the case when the constraint is
strictly imposed.

The constraint (\ref{cons3}) requires the low energy Hamiltonian
to be invariant under the gauge transformation $A_{ij} \rightarrow
A_{ij} + \partial_if_j + \partial_jf_i$. This is precisely the
gauge symmetry of the graviton in 3 dimensional space. Thus, the
low energy physics can only involve the linearized curvature
tensor. In the 3 dimensional space, the curvature tensor has 6
nonzero components, according to the symmetry of the curvature
tensor. Thus, now the effective low energy Hamiltonian reads \beqn
H_{ring} = \sum_{ij,i\neq j}-\tilde{t}_1\cos(R_{ijij}) -
\sum_{ijk,i\neq j, j\neq k, i\neq k} \tilde{t}_2\cos(R_{ijik}) \cr
+ \frac{1}{2\kappa_1}(\sum_{i = 1}^3E^2_{ii}) +
\frac{1}{2\kappa_2}(\sum_{ij,i\neq j}E_{ij}^2).\cr
\label{ring}\eeqn The cosine terms involving the curvature tensor
are ring exchange terms generated by the nearest neighbor hopping.
Ring exchanges in (\ref{ring}) are generated at the eighth order
perturbation of the nearest neighbor hopping, $\tilde{t}_1$,
$\tilde{t}_2\sim t^8/V^7$. All the lower order perturbations only
generate terms which do not comply with the constraint
(\ref{cons3}). The ring exchange term $\cos(R_{xyxy})$ is the same
as its 2 dimensional counterpart, as shown in Fig. \ref{2dring}.

In order to derive the correct Lagrangian, one needs to introduce
a Lagrange multiplier $A_{i0}$ for the constraint (\ref{cons3}).
The full Lagrangian reads \beqn L =
\sum_{i,j}(\partial_{\tau}A_{ij} -
\partial_iA_{j0} - \partial_{j}A_{i0})^2 +
\sum_{ij,i\neq j}\tilde{t}_1\cos(R_{ijij}) \cr + \sum_{ijk,i\neq
j, j\neq k, i\neq k} \tilde{t}_2\cos(R_{ijik}) \cr
\label{3daction}\eeqn The gauge symmetry can now be enlarged to
quantities defined in the 3+1 dimensional space-time: $A_{\mu\nu}
\rightarrow A_{\mu\nu} +
\partial_\mu f_\nu + \partial_\nu f_\mu $ and $A_{00} = 0$, $f_{0} =
0$. In this system, the boson superfluid order is again ruled out
by the gauge symmetry. Without crystalline order (proven later),
the system is in a liquid phase with excitations which have the
same gauge symmetry as the graviton. In the linearized Einstein
gravity, after taking the traceless-transverse gauge, the spin-2
graviton has only two polarizations. In our theory tracelessness
was not imposed to $A_{ij}$. Therefore there are three gapless
collective excitations in the graviton phase, one of which is the
scalar trace mode, the other two are described by traceless
matrices, if the transverse gauge is taken, the two traceless
modes exactly correspond to the two polarizations of the
gravitons.

Unlike the quantum dimer model, the curvature tensor is the second
spatial derivative of $A_{ij}$. If there is a Gaussian phase in
which we can expand the cosines in equation (\ref{3daction}), the
gapless graviton mode in this Gaussian phase has a soft dispersion
$\omega \sim k^2$.

Whether the graviton excitations survive (or equivalently whether
crystal order develops) can be studied in the dual theory. If we
define the symmetric tensor $\mathcal{E}_{ij}$ as
$\mathcal{E}_{ii} = 2E_{ii}$, $\mathcal{E}_{ij} = E_{ij}, i \neq
j$, the constraint (\ref{cons3}) can be solved by defining the
dual tensor $h_{ij}$ as \beqn \mathcal{E}_{ij} =
\epsilon_{iab}\epsilon_{jcd}\partial_a\partial_ch_{bd}. \eeqn This
is a double curl of the symmetric tensor $h_{ij}$. $h_{ij}$ also
lives on the sites and faces of this fcc lattice.

If checking carefully, one can notice that, the curvature tensor
can also be written in the double curl form  \beqn 2R_{xyxy} =
\epsilon_{zab}\epsilon_{zcd}\partial_a\partial_cA_{bd}, 2R_{xzxz}
= \epsilon_{yab}\epsilon_{ycd}\partial_a\partial_cA_{bd} ,\cr\cr
2R_{yzyz} =
\epsilon_{xab}\epsilon_{xcd}\partial_a\partial_cA_{bd}, 2R_{xyxz}
= \epsilon_{yab}\epsilon_{zcd}\partial_a\partial_cA_{bd},\cr\cr
2R_{yxyz} =
\epsilon_{xab}\epsilon_{zcd}\partial_a\partial_cA_{bd}, 2R_{zxzy}
= \epsilon_{xab}\epsilon_{ycd}\partial_a\partial_cA_{bd}. \cr
\label{doublecurl}\eeqn Therefore, this model is self-dual, as
long as we define the dual variables $h_{ij}$ in terms of $
\mathcal{E}_{ij} =
\epsilon_{iab}\epsilon_{jcd}\partial_a\partial_ch_{bd}$ and its
conjugate $\pi_{ij}$ as follows: \beqn R_{xyxy} = \pi_{zz},
R_{yzyz} = \pi_{xx}, \cr\cr R_{xzxz} = \pi_{yy}, 2R_{xzyz} =
\pi_{xy}, \cr\cr 2R_{xyxz} = \pi_{yz}, 2R_{xyzy} = \pi_{xz}.
\label{dualpi}\eeqn According to the definition, $\pi_{ij}$ is
subject to the same constraint as $E_{ij}$ (\ref{cons3}).

After introducing the dual variables $h_{ij}$ and $\pi_{ij}$, the
dual Lagrangian reads \beqn L_{dual} = \sum_{ij}(\partial_th_{ij}
-
\partial_ih_{j0} -
\partial_jh_{i0})^2 - \sum_{ij,i\neq j}\rho_1\tilde{R}^2_{ijij}
\cr - \sum_{ijk,i\neq j,j\neq k, i\neq k}\rho_2\tilde{R}^2_{ijik}
+ \cdots\cr \label{dualac3}\eeqn $\tilde{R}_{ijkl}$ is the
curvature tensor of $h_{ij}$. $h_{i0}$ is a Lagrange multiplier,
which is introduced for the constraint on $\pi_{ij}$. The ellipses
include possible vertex operators. Without the vertex operators,
this theory is at a Gaussian fixed point and hence in an algebraic
liquid phase with soft graviton excitations. If the vertex
operators are relevant, they will destabilize the liquid phase and
gap out the graviton excitation, and form crystalline order
according to the Berry phase. The dual Lagrangian (\ref{dualac3})
is also invariant under the gauge transformation $h_{\mu\nu}
\rightarrow h_{\mu\nu} +
\partial_\mu f_\nu +
\partial_\nu f_\mu$, with $h_{00} = 0$ and $f_0 = 0$.
Therefore, any kind of vertex operator (for example $\cos(2N\pi
h_{ij})$) is not a gauge invariant operator. The correlation
function between two vertex operators at the Gaussian fixed point
is zero or correlated at very short range (the correlation length
is supposed to be roughly the inverse of the charge gap $V$ in
$H_0$ (\ref{h0})), so the Gaussian fixed point (also the algebraic
spin liquid) is stable against weak perturbations of the vertex
operators. Thus, this graviton phase is stable for the same reason
as the photon phase, as discussed in the second section of this
paper.

\section{properties of the algebraic liquid phase}

The most important feature of the algebraic spin liquid phase is
the power law correlation between spin operators. In the boson
language used in this paper, it is the boson density operators
which correlate algebraically. As discussed before, in the
algebraic liquid phase, the boson density fluctuation around the
average filling $\bar{n}$ can be written as (for instance)
$n_{1,i} - \bar{n} \sim \eta_i
\epsilon_{1ab}\epsilon_{1cd}\partial_a\partial_ch_{bd}$. The
calculation of the equal time boson density correlation can be
derived from the correlation functions between $h_{ij}$. \beqn
\langle (n_{1,i} - \bar{n})(n_{1,j} - \bar{n}) \rangle \sim
\eta_i\eta_j
\partial^2_i\partial^2_j\langle h(i)h(j) \rangle \cr\cr \sim \eta_i\eta_j
\partial^4 (\frac{1}{r}) \sim \eta_i\eta_j\frac{1}{r^5}. \eeqn $r = |i -
j|$. Therefore the correlation functions between boson density
operators fall off algebraically, with an exponent higher than the
exponent of the photon liquid phase.

Because of the gauge symmetry, the operators have to be gauge
invariant to have nonzero correlation functions. Therefore the
correlation function between original boson creation operators
$b_i$ are zero (or shortly correlated with a correlation length
which roughly equals to the inverse of the charge gap $V$). The
curvature tensors have nonzero correlations, and the correlators
also fall off algebraically with the same exponent as that of the
density correlator.

The dynamics of the gapless liquid phase can be described by a new
set of Maxwell's equations. Define the rank-2 tensor
$\mathcal{B}_{ij} =
\epsilon_{iab}\epsilon_{jcd}\partial_a\partial_c A_{bd}$, the
dynamical equations that describe this liquid phase can be derived
directly from the Lagrangian (\ref{3daction}), \beqn
\partial_i\mathcal{E}_{ij} = 0, \cr\cr \partial_i\mathcal{B}_{ij} = 0,\cr
\cr \partial_t\mathcal{E}_{ij} -
\kappa\epsilon_{iab}\epsilon_{jcd}
\partial_a\partial_c\mathcal{B}_{bd} = 0,\cr\cr
\partial_t\mathcal{B}_{ij} + \kappa\epsilon_{iab}\epsilon_{jcd}
\partial_a\partial_c\mathcal{E}_{bd} = 0.\label{maxwell}\eeqn Charged excitations
and topological defects are absent in these equations, thus the
equations correspond to the Maxwell's equations in vacuum. If the
constraint (\ref{cons3}) is softened, charge density and charge
current have to be incorporated in equation (\ref{maxwell}).

\begin{figure}
\includegraphics[width=2.5in]{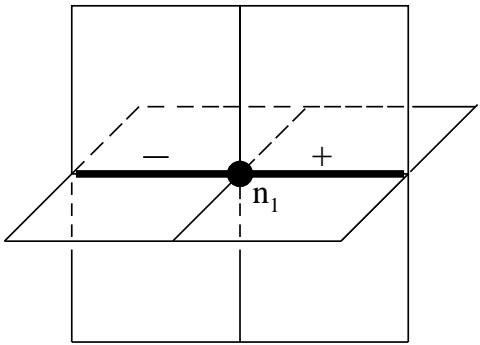}
\caption{one of the defects carrying gauge charges. If $n_1$ on
one site is increased by one, it is equivalent to creating a pair
of opposite gauge charges on two $\hat{x}$ links sharing this
site.} \label{defect}
\end{figure}

\begin{figure}
\includegraphics[width=2.5in]{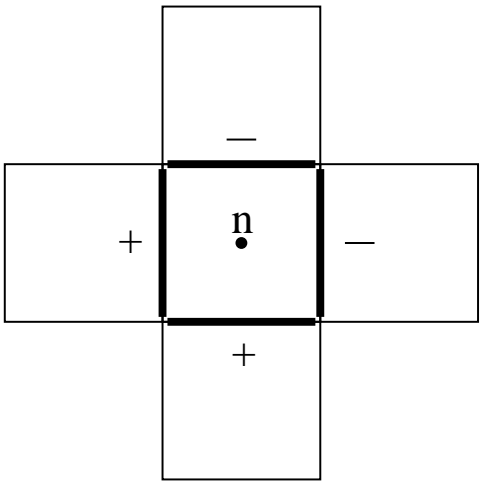}
\caption{one of the defects carrying gauge charges. if on one
plaquette center, $n_{\bar{r}}$ is increased by one, gauge charges
are created for four links sharing this plaquette.}
\label{defect2}
\end{figure}

The violation of constraint (\ref{cons3}) corresponds to the
defects, which carry gauge charges of gauge fields $A_{ij}$.
Because the gauge field is a rank-2 tensor, the gauge charge
should be vector field. The static vector charge field couples to
the gauge field in a similar way with the Gauss's law: $
\partial_i\mathcal{E}_{ij} = -\rho_j $. For instance, if on one
site, we increase $n_1$ by one, it is equivalent to excite a pair
of opposite gauge charges on the two $\hat{x}$ links sharing this
site (Fig. \ref{defect}); if on one plaquette center,
$n_{\bar{r}}$ is increased by one, gauge charges are created for
four links sharing this plaquette (Fig. \ref{defect2}). Notice
that since the constraint imposed by $H_0$ in the original
Hamiltonian (\ref{h0}) is an interaction between particles around
each link, now the gauge charge field is defined on links.

The defects should couple to the gauge field in a gauge invariant
manner, in the rotor language, the coupling can be written as
\beqn H_e = - \sum_{i,j} t\cos(\partial_i\theta^{(e)}_j +
\partial_j\theta^{(e)}_i - A_{ij}) + \cdots \eeqn  $\theta^{(e)}_i$
are the phase angles of gauge charge field creation operators.
This coupling is gauge invariant since if $\theta^{(e)}_i$ is
added by any arbitrary function $f_i$, one can always eliminate
this extra phase angle by gauge transformation $A_{ij}\rightarrow
A_{ij} + \partial_if_j +
\partial_jf_i$. In the Gaussian phase, the defects are gapped, and
the phase angles $\theta^{(e)}_i$ are in the disordered phase. If
the defects condense, i.e. the phase angles $\theta^{(e)}_i$ take
nonzero expectation values, the $A_{ij}$ will be gapped out
through Higgs mechanism. However, the matter field phason modes
will not be completely gapped out. This is because of the extra
global total particle number conservation besides the gauge charge
conservation. After the condensation of all the matter fields, the
gauge charge conservation will be broken, and the gauge field is
gapped out through the Higgs mechanism. Meanwhile, the spontaneous
breaking of the global particle number conservation guarantees the
existence of the gapless excitation in the condensate, which is
exactly the Goldstone mode. Notice that, although there are in
total four flavors of matter fields ($n_1$, $n_2$, $n_3$ and $n$),
different flavors of matter fields are mixed in the nearest
hopping term $H_1$ (\ref{3dh1}), therefore only the total boson
number is conserved. There should be only one Goldstone mode in
the condensate. Similar situation takes place in the doped quantum
dimer model \cite{balents2005b}, where the holes carry both the
gauge charge and the global $U(1)$ charge, hence the condensate of
holes contains one gapless Goldstone mode.

In the dual formalism, the topological defects can be introduced
in the vertex operators of the dual formalism. The topological
defects can be introduced in the dual vertex operators \beqn H_m =
- \sum_{i,j} t\cos(2\pi N(\partial_i\theta_j^{(m)} +
\partial_j\theta_i^{(m)}) - 2\pi Nh_{ij}) \cr \eeqn
$\theta^{(m)}_i$ are the phase angles of the creation operators of
the topological defects. One can see the topological defects
couple to the dual gauge potential $h_{ij}$ in the same manner as
the gauge charges couple to the original gauge potential $A_{ij}$,
i.e. the charge and topological charge are dual to each other,
which is the same case as the duality between the electric charge
and the magnetic monopole. After the condensation of the
topological defects $\theta^{(m)}_i$, the gapless graviton
excitation $h_{ij}$ is gapped out by Higgs mechanism. However, in
this situation, there is no extra global conservation of the
topological defects, therefore in the condensate there will not be
any gapless Goldstone mode. The system is in the gapped Mott
Insulator phase, with crystalline order determined by the Berry
phase. Since our emphasis of this paper is about the stable liquid
phase, we will not get into the detailed analysis of the
crystalline phase.

After introducing the matter field and the topological defect, the
semiclassical Maxwell's equations with charges are

\beqn
\partial_i\mathcal{E}_{ij} = - \rho^{(e)}_j, \cr\cr
\partial_i\mathcal{B}_{ij} = - \rho^{(m)}_j,\cr
\cr \partial_t\mathcal{E}_{ij} -
\kappa\epsilon_{iab}\epsilon_{jcd}
\partial_a\partial_c\mathcal{B}_{bd} = J^{(e)i}_j + J^{(e)j}_i,\cr\cr
\partial_t\mathcal{B}_{ij} + \kappa\epsilon_{iab}\epsilon_{jcd}
\partial_a\partial_c\mathcal{E}_{bd} = J^{(m)i}_j + J^{(m)j}_i.\label{maxwellc}\eeqn

The current $J^{(e)i}_j$ represents the current with the $i$th
component of matter field flowing in $\hat{j}$ direction. The
conservation law of the gauge charges, as well as the topological
defects reads \beqn
\partial_t\rho^{(e)}_j = - \partial_iJ^{(e)j}_i - \partial_iJ^{(e)i}_j,\cr\cr
\partial_t\rho^{(m)}_j = - \partial_iJ^{(m)j}_i - \partial_iJ^{(m)i}_j. \eeqn

One can clearly see that, because of the flavor mixing in $H_1$,
the vector charge density is not individually conserved. The
derivative $\partial_iJ^{(e)i}_j$ plays the role of torque density
of the vector charge density. This is similar to the spin current
\cite{zhang2003}, which is also a tensor current, and the
conservation equation of the spin current involves torque density
\cite{niu2004}, due to the fact that spin is not conserved in a
system with spin-orbit coupling.

\section{conclusions, experimental realizations and extensions}

In this work we constructed models which give rise to collective
excitations analogous to the gravitons. In 2+1 dimensional system,
the graviton excitations are unstable against vertex operators in
the dual formalism; in 3+1 dimensional space, there is a stable
algebraic boson liquid phase which contains gapless graviton
excitations with soft dispersion. The graviton liquid phase is
self-dual, and the stability of the algebraic phase is guaranteed
by the large gauge symmetry on both sides of the duality. The
dynamics of the phase is described by a new set of Maxwell's
equations.

We proved the algebraic phase is stable, i.e. the realization of
this phase does not require any fine-tuning. However, the original
bosonic model is of a special form, its precise realization in
experimental systems requires more efforts. Here we only consider
the possibility of its realization in the cold atom system trapped
in an optical lattice. The original boson model contains standard
nearest neighbor hopping between particles, and also contains two
body interactions, both on-site and off-site. The on-site
repulsion can be obtained from the $s$-wave scattering between
bosons on the same site, and the off-site repulsion has been shown
recently to exist in Chromium atom condensate due to the long
range dipole interaction \cite{axel2005,stuhler2005}.

The interaction between the three orbital levels on each site and
the bosons in the plaquette centers is very anisotropic
(\ref{h0}), this could be due to the anisotropy of the orbital
level spatial wave functions. In the transition metal oxides
materials, the anisotropy of the $t_{2g}$ level electron wave
functions gives rise to the Khaliullin model
\cite{khaliullin2000,khaliullin2002} which was supposed to explain
the orbital liquid phenomenon, in which the orbital moment is
quenched by quantum fluctuation \cite{keimer2000}. Recently it has
been proposed that, although the ground state of each site of the
optical lattice is $s$-wave, if the particles are pumped to the
first excited three fold degenerate $p$-wave states, the particles
will maintain in the excited states for a considerable time, long
enough for the particles to equilibrate \cite{girvin2005},
therefore the $p$-wave particles will first form equilibrium state
before they drop to the $s$-wave ground states. Each of the three
fold degenerate $p$-level wave functions only extends in one
direction of the 3 dimensional space. Since the wave function
overlap is anisotropic in space, the particle on each orbital
level interacts more strongly in one certain direction, as long as
the optical trap on each site is not too deep, i.e. the wave
function is not too localized on each site. The anisotropic
interaction is required in our problem. The remained problem of
the experimental realization is how to realize the fcc lattice
with laser beams, and how to pump particles on all the sites of
the fcc lattice to the excited $p$-wave states.

So far the models we have considered are purely bosonic models.
After the search for unconventional bosonic phases, one might
wonder if non-fermi liquid can be obtained in a similar way for
fermionic systems. There has been a great deal of study on
treating the constraint of no double occupancy of Hubbard model by
introducing $U(1)$ gauge field, and the bosonic holons and
fermionic spinons are interacting with this $U(1)$ gauge field.
The interacting system has been proposed to explain the nonfermi
liquid behavior of the normal state of High Tc cuperates (for
instance, reference \cite{patrick1993}). In this kind of theories,
the emergent $U(1)$ gauge field plays the most crucial role. For
instance, due to the scattering from the gauge fields, the scaling
of the resistivity significantly deviates from the $T^2$ law of
normal fermi liquid. The emergence of this gauge field is exactly
due to the local constraint $\sum_{\sigma} f^\dagger_{i,\sigma}
f_{i,\sigma} + b^\dagger_ib_i = 1$.

In our current work, if all the particles in the original model
are fermions, the constraint imposed by $H_0$ can also be solved
by introducing rank-2 tensor gauge field $A_{ij}$, and when the
gauge field is in its deconfined phase, the system can be viewed
as fermions interacting with gapless soft graviton mode $A_{ij}$.
The behavior of the fermions is expected to be non-fermi liquid.

In the whole paper, our goal has been carefully limited to the
discovery of a new type of algebraic spin liquid. However, one can
consider the issue of "quantum gravity" as an extension of this
work. Indeed, if in a theory the gauge symmetry of the graviton
can emerge at low energy physics, the theory might be a candidate
of a new possibility of quantum gravity. Several other systems in
condensed matter physics have been proposed to be related to
gravity, for instance, spin-2 particles are supposed to exist at
the edge states of 4 dimensional quantum Hall model
\cite{zhang2001}. However, in these systems the gauge symmetry
(which is very crucial for gravitons) was not an emergent
property. In our work, the collective excitations automatically
have the gauge symmetry of the graviton. However, the dispersion
relation is quadratic, this is due to the fact that the gauge
invariant operator, the curvature tensor is the second derivative
of $A_{ij}$. The ring exchange term generated from the nearest
neighbor hoppings is $\cos(R)$, after the expansion, the leading
term is $R^2$, which is proportional to $k^4$.

We can make the graviton dispersion linear by introducing by hand
a quasi-gauge invariant term to the low energy Lagrangian : \beqn
L_{cs} =
A_{ij}\epsilon_{iab}\epsilon_{jcd}\partial_a\partial_cA_{bd}.
\label{cs}\eeqn This term is not completely gauge invariant,
instead, it is gauge invariant up to a boundary term. This is very
analogous to the Chern-Simons field theoy of the Quantum Hall
effect \cite{zhang1989}. However, this term cannot be generated
from the microscopic boson model with only nearest neighbor
hopping. It is expected that, by coupling to the matter fields,
and the matter fields form certain special state, the $k^2$ term
(\ref{cs}) can be generated from integrating over the matter
fields in the partition function, just like how the Chern-Simons
field theory is obtained in the Quantum Hall state of electrons.
Recently, some other authors have proposed a bosonic model which
is similar to ours, and the $k^2$ term (\ref{cs}) is claimed to
exist in the long scale physics, therefore a graviton with linear
dispersion is supposed to exist \cite{wen2006}.

\bibliography{longravity}
\end{document}